\documentclass[conference]{IEEEtran}
\IEEEoverridecommandlockouts
\usepackage{cite}
\usepackage{amsmath,amssymb,amsfonts}
\usepackage{algorithmic}
\usepackage{graphicx}
\usepackage{textcomp}
\usepackage{xcolor}

\usepackage{color} 
\usepackage{bm}
\usepackage{subfigure}
\usepackage{epstopdf}
\usepackage{cite}
\usepackage{graphicx,algorithm}
\usepackage{makecell}
\usepackage{color}   
\usepackage{array}
\usepackage{multicol}
\usepackage{flushend, cuted}
\usepackage{comment}
\usepackage{multirow}
\usepackage{longtable}
\allowdisplaybreaks

\def\BibTeX{{\rm B\kern-.05em{\sc i\kern-.025em b}\kern-.08em
    T\kern-.1667em\lower.7ex\hbox{E}\kern-.125emX}}

\begin{document}

\title{Characterization of Wireless Channel Semantics: \\A New Paradigm\\
}
\DeclareRobustCommand*{\IEEEauthorrefmark}[1]{%
  \raisebox{0pt}[0pt][0pt]{\textsuperscript{\footnotesize\ensuremath{#1}}}}
    \author{\IEEEauthorblockN{Zhengyu Zhang\IEEEauthorrefmark{\dag}, Ruisi He\IEEEauthorrefmark{\dag}, Mi Yang\IEEEauthorrefmark{\dag}, Xuejian Zhang\IEEEauthorrefmark{\dag}, Ziyi Qi\IEEEauthorrefmark{\dag}, Yuan Yuan\IEEEauthorrefmark{\dag} and Bo Ai\IEEEauthorrefmark{\dag}}

\IEEEauthorblockA{\IEEEauthorrefmark{\dag}School of Electronics and Information Engineering, Beijing Jiaotong University, Beijing 100044, China}

Ruisi He (Corresponding Author). e-mail: ruisi.he@bjtu.edu.cn.

}
\maketitle

\begin{abstract}

Recently, deep learning enabled semantic communications have been developed to understand transmission content from semantic level, which realize effective and accurate information transfer. Aiming to the vision of sixth generation (6G) networks, wireless devices are expected to have native perception and intelligent capabilities, which associate wireless channel with surrounding environments from physical propagation dimension to semantic information dimension. Inspired by these, we aim to provide a new paradigm on wireless channel from semantic level. A channel semantic model and its characterization framework are proposed in this paper. Specifically, a channel semantic model composes of status semantics, behavior semantics and event semantics. Based on actual channel measurement at 28 GHz, as well as multi-mode data, example results of channel semantic characterization are provided and analyzed, which exhibits reasonable and interpretable semantic information.

\end{abstract}


\section{Introduction}
The development of wireless communication systems has led to the emergence of the sixth generation (6G) network, considering as a revolutionary shift from traditional communication paradigms \cite{1}. More than just an improvement of existing communication technologies, 6G is characterized by its potential to provide ubiquitous sensing and connectivity. In this vision, integrated sensing and communication (ISAC) techniques are encouraged to enhance wireless devices with native perceptual capabilities, making it possible to “see the environment” for wireless communication systems \cite{2}. Besides, based on the development of artificial intelligence (AI) and computer vision (CV), it is empowered for wireless communication systems to see the environment more deeply, i.e., to “understand the environment”, achieving improved communication performance and intelligence decision with adaption to dynamic environment. Lots of emerging applications have been prompted under detailed semantics information, for instance, environment reconstruction, enhanced positioning and tracking, human pose detection, vehicle detection and so on \cite{3}. Semantic information has become a novel and potential information dimension for 6G networks.

Accurate semantic information can provide structured knowledge networks and scalable knowledge library for information source, channel, and task accordingly, which supports the efficient operation and updates of intelligent communication agents. For example, semantic communication, which attracted widespread attention in recent years, has been conducted semantic aware joint source and channel coding for effective and accurate information transfer based on semantic knowledge library \cite{4,5,6}. However, semantic communication focuses on the transmission of content from transmitter to receiver, which usually includes redundant information in physical signals compared to represented meanings. And in semantic communication, semantic channel is considered as the pipeline for semantic transmission, which poses challenges in connecting with the surrounding environment.

In order to associate the wireless channel with surrounding environmental information, channel semantics are gradually gaining attention to provide a representation for environmental and behavioral information, particularly in mapping with channels and scatterers. Ref \cite{7} present the results of wave scatterer localization at 4.65 and 14.25 GHz based on a single-bounce channel model. Ref \cite{8} provides a study of sensing assisted environment reconstruction and communication in ISAC scenario. The radio-based sensing and environment mapping on the user equipment side is provided in \cite{9}. However, these researches conduct environment mapping or reconstruction without understandable semantic information. In fact, wireless channel semantics is potential to provide a highly efficient environment perception and understanding for communication systems, assisting in prediction, decision, beam forming, and so on. Although some semantic information is designed for a specific task or communication system, for example, Ref \cite{10} defined task-oriented propagation environment semantics, and Ref \cite{11} proposed a predictive channel-based semantic communication system tailored for sensing scenarios. Further exploration and generalized research about channel semantics are needed to uncover semantic information related to the environment, which directly reveals radio propagation from transmitter to receiver, rather than redundant environment reconstruction.

\begin{figure*}[!t] \center \vspace{0in}
\includegraphics[width=1\textwidth]{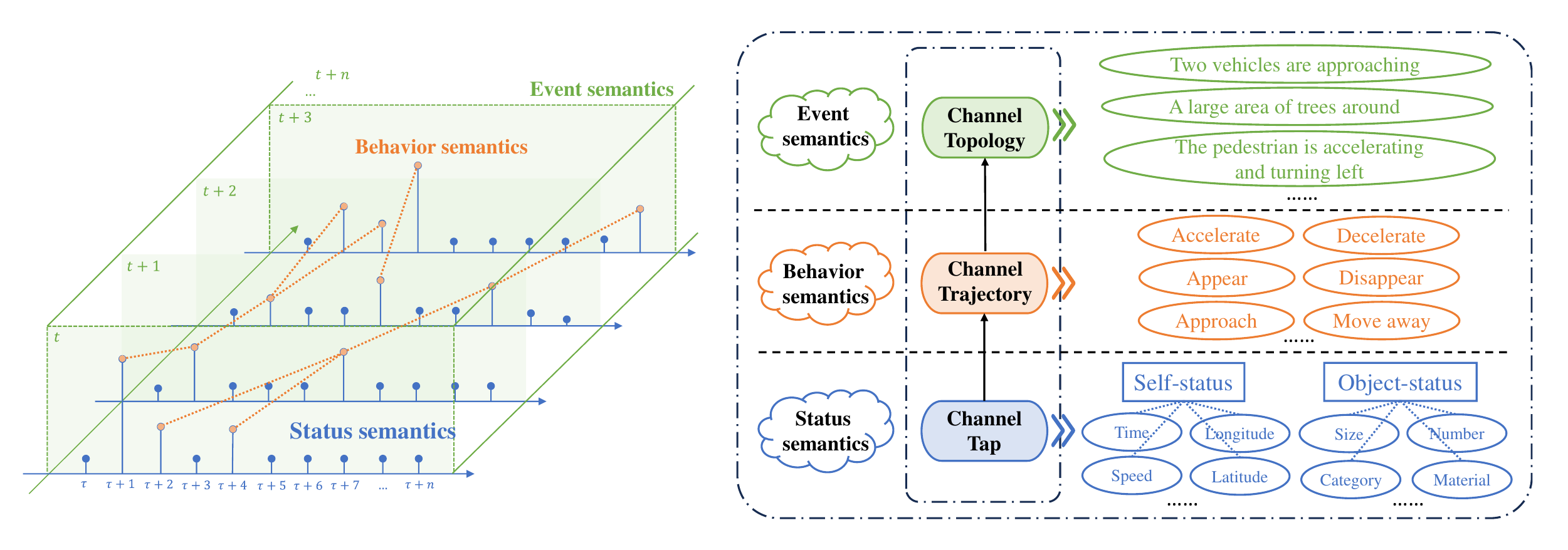}\vspace{-0in}
\caption{Channel semantic model with multi-level semantic information.
\label{fig:model}}\vspace{-0.1in}
\end{figure*}

In this paper, a channel semantic model and its characterization framework is proposed. Specifically, a channel semantic model composes of status semantics, behavior semantics and event semantics, corresponding to channel taps, channel trajectories and channel topologies respectively. Based on actual measurement, the channel semantic characterization results are provided and analyzed. The research in this paper aims to provide a new paradigm on wireless channel at semantic level, which enable efficient environment perception and understanding for communication system.

The remainder of this paper is organized as follows. Section II introduces the semantic channel model in details. Section III describes the semantics characterization based on channel measurement. Section IV presents and analyzes the results of channel semantic characterization. Finally, Section V draws the conclusions

\section{Channel Semantic Model}

As shown in Fig. 1, a channel semantic model is proposed to decouple tradition time-varying channel models as multi-level semantic information. In the proposed model, the channel semantics consist of status semantics, behavior semantics and event semantics, which are used to construct the semantic-level wireless propagation environment. To meet the interpretability and scalability, the channel semantic model should

$\bullet$ express the concepts and the implicit correlations between status and behaviors under typical event scenario;

$\bullet$ store the historical behaviors for reasoning, trajectory annotation, and semantic query;

$\bullet$ contain the potential and reasoning status from the basic behavior.

The channel semantic model is in the form of triple as follows:
\begin{align}\label{BS_RIS}
CS = \{\mathbf{E}, \mathbf{B}, \mathbf{S}\}
\end{align}
where $CS$ represents the channel semantics; $\mathbf{E}=\{E_1, E_2,..., E_n\}$ denotes the $\mathit{Event}$ (all semantic terms in the channel semantic model are italic in the paper), which contains the multiple event concepts over a period. It is noted that these events can be correlated or independent, occurring simultaneously or sequentially. $\mathbf{B}=\{B_1, B_2,..., B_n\}$ denotes the $\mathit{Behavior}$, which is a collection of a set of behaviors within an $\mathit{Event}$. Only a set of behaviors that jointly impact a specific event are considered as a $\mathit{Behavior}$. $\mathbf{S}=\{S_1, S_2,..., S_n\}$ denotes the $\mathit{Status}$, which is the description of a specific object, and corresponds to one or multiple behaviors. A more detailed introduction is provided in the following text.

\subsection{Event Semantics}

$\mathit{Event}$, which represent the macroscopic behavior relative to the current observation view, describe the logical and topological relationships between the behaviors. For a specific event, $E_n$ is defined as follows:
\begin{align}\label{BS_RIS}
E_n = \{t_n, T_n, B_{i\in \mathbb{M}}\}
\end{align}
where $t_n$ is the start time of event $E_n$, $T_n$ is the duration of event $E_n$, $B_{i\in \mathbb{M}}$ is the set \(\mathbb{M}\) of combination behaviors, which correspond to event $E_n$. For example, the entire process of two vehicles approaching can be considered as one event. $\mathit{Event}$ can also denote the process that occurs in a larger temporal and spatial range. For example, the entire process of vehicles stopping, as well as pedestrians accelerating across the road, can be considered as an event “intersection cycle”. The event corresponding to the multiple behaviors can represent a semantic aggregation with related relationships. 

\subsection{Behavior Semantics}

$\mathit{Behavior}$, which represent the action behavior over continuous time variation, occurs along the channel trajectory within a range of delay and time. For a specific behavior, $B_n$ is defined as follows:
\begin{align}\label{BS_RIS}
B_n = \{t_n, T_n, \tau_n, D_n, S_{i\in \mathbb{N}}\}
\end{align}
where $t_n$ is the start time of behavior $B_n$, $T_n$ is the duration of behavior $B_n$, $\tau_n$ is the start location in delay of behavior $B_n$, $D_n$ is the coverage in delay of behavior $B_n$, $S_{i\in \mathbb{N}}$ is the set \(\mathbb{N}\) of combination status, which correspond to behavior $B_n$. Different from events, behavior only reflects the semantics of motion characteristics of the trajectory, such as accelerate and decelerate, appear and disappear, approach and move away and so on. Therefore, it does not include the semantics of the interaction between the trajectory and its surroundings. In contrast, events contain more semantic information due to topological relationships. 

\subsection{Status Semantics}

$\mathit{Status}$, which represent the immediate semantic features relative to the single snapshot, describe the self-characteristics or object-characteristics with channel delay taps respectively. For a specific status, $S_n$ is defined as follows:

\begin{align}\label{BS_RIS}
S_n = \{\tau_{i\in \mathbb{R}}, A_{i\in \mathbb{R}}\}
\end{align}
where $\tau_{i\in \mathbb{R}}$ and $A_{i\in \mathbb{R}}$ are the set \(\mathbb{R}\) of delay and amplitude respectively, which correspond to status $S_n$. 
Status semantics are utilized to describe the fundamental environment without incorporating additional semantic information related to time. For example, self-status semantics, which are affected by wireless device itself, include time, speed, location (such as longitude and latitude) and so on. Object-status semantics, which are affected by surrounding scatterers and perception targets, include size, number, category, material and so on. As the basic semantic units in channel semantics model, the continuous status semantics, enriched with behavior semantics, can be constructed based on the channel trajectory.

\section{Semantics characterization based on channel measurement}

In this section, we present semantics characterization based on actual channel measurements as examples, to clearly show spirit of channel semantics.

\subsection{Channel Measurement Campaign}

\begin{figure}[htbp]
\centering
\subfigure[]{\includegraphics[width=2in]{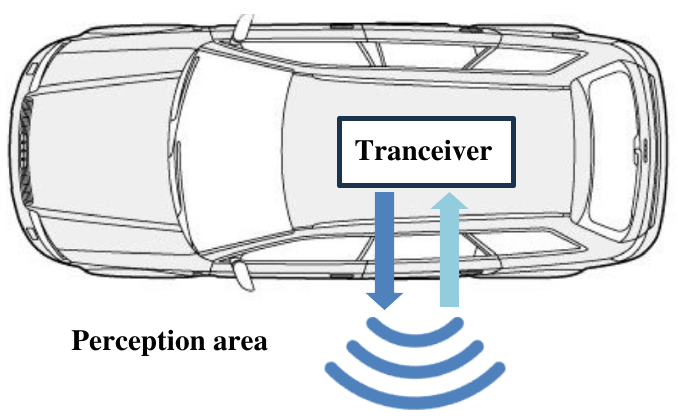}}
\subfigure[]{\includegraphics[width=1.5in]{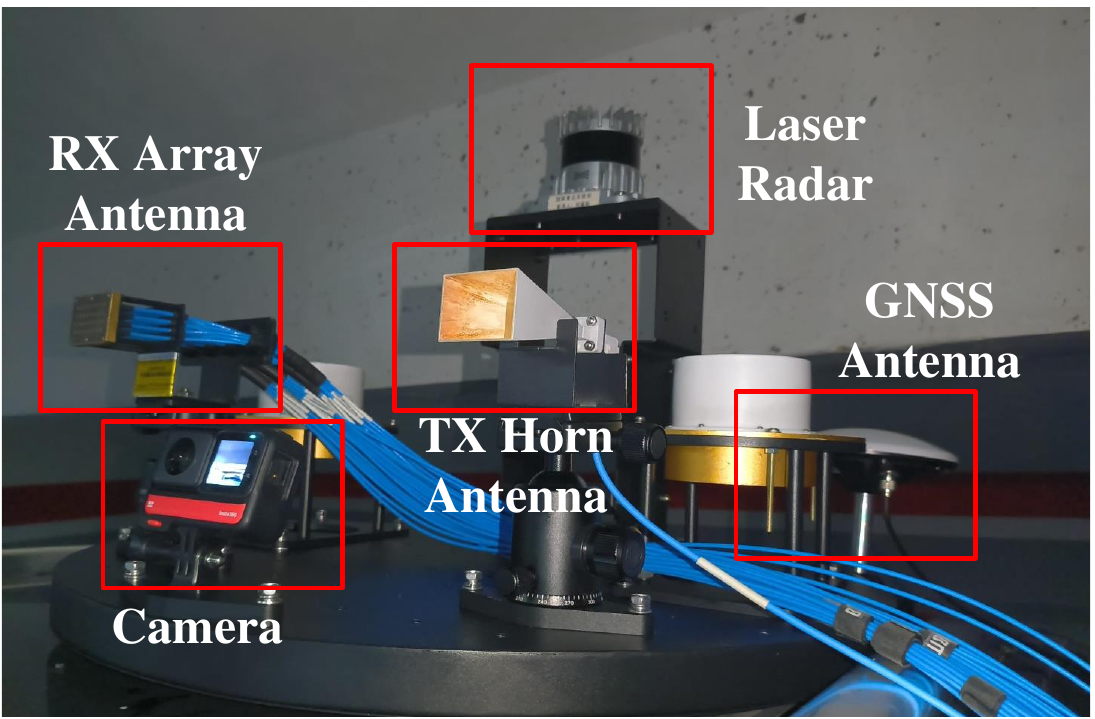}}
\subfigure[]{\includegraphics[width=1.5in]{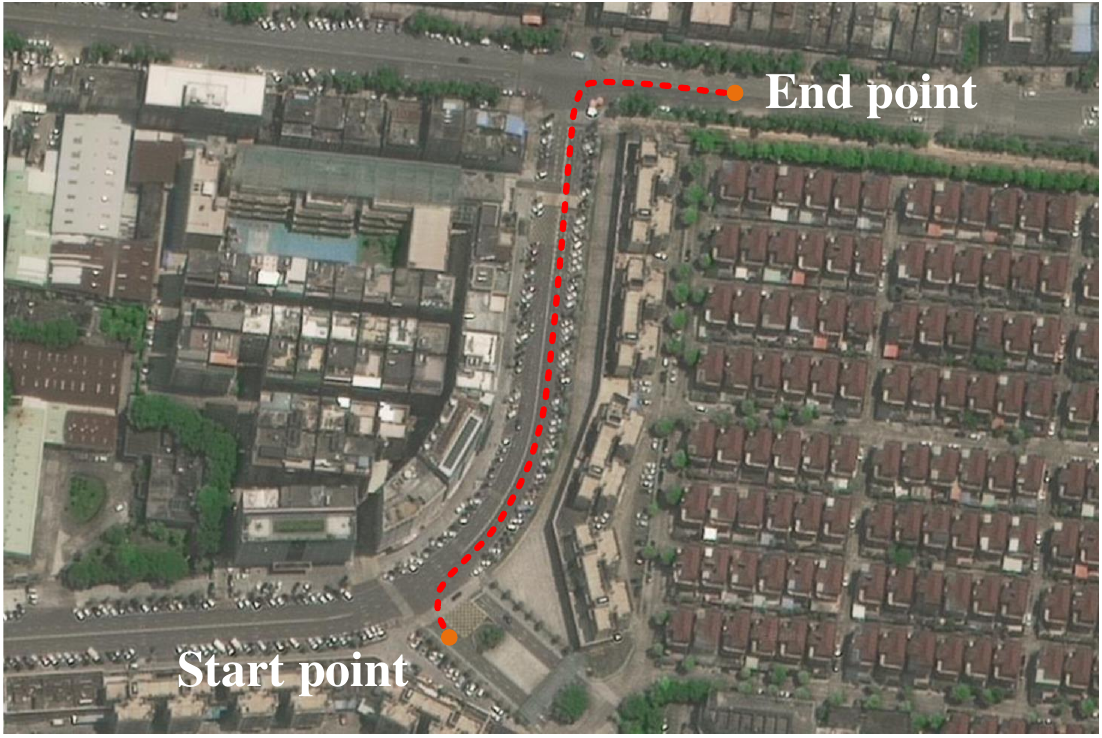}}
\caption{Channel measurement campaign. (a) layout. (b) equipment. (c) scenario.}
\label{fig}
\end{figure}

In order to analyze comprehensive environmental information, the vehicular multi-mode channel measurements, including point-cloud data, RGB data, and channel data, are conducted at 28 GHz in urban scenarios. The channel measurement layout, equipment and scenario are illustrated in Fig. 2, which is designed for ISAC scenarios \cite{13}. Considering the distribution and variation of scatterers, the perception area is located on the left side of the vehicles, as shown in Fig. 2(a). The measurement equipments, which includes laser radar, camera, TX hoen antenna, RX array antenna and GNSS antenna, are mounted on vehicles with a height of 2.3 m, and other controller equipment is inside the vehicles. The camera and laser radar both have a 360 degree vision. The TX horn antenna has a 18 degree beamwidth with 20 dB gain, and RX array antenna is a 32-element rectangular array with 5 dB gain. At TX, the signal generator is employed to generate the baseband sounding signals with 1 GHz bandwidth, and millimeter wave RF transmission module upconvert signals to 28 GHz. At RX, sounding signals undergo downconversion to baseband through millimeter wave RF receiving module. And they are stored by signal digitizer and storage device. The synchronization of the transmitter and receiver is achieved through rubidium clocks and GNSS antennas, ensuring a 10 MHz reference clock. The sounding signals are a group of sinusoidal signals with different frequency points in the measured bandwidth, and are transmitted through a power amplifier with an output of 28 dBm.

The measurements were conducted in the Songshanhu district of Dongguan, located in Guangdong, China. The measurement-related street is urban scenarios and the route is shown in Fig. 2(c), which is surrounded by densely arranged buildings. During the measurements, the mean speed of vehicle is 20 km/h and the duration is 60 s. The main scatterers are buildings, trees, vehicles and median barriers. In the measurements, the whole process is considered as an event “driving through road”, and more detailed events and behaviors are considered as “turn onto road”, “yield to other vehicles”, “go straight” and so on.

\subsection{Characterization Framework}

\begin{figure*}[!t] \center \vspace{0in}
\includegraphics[width=0.95\textwidth]{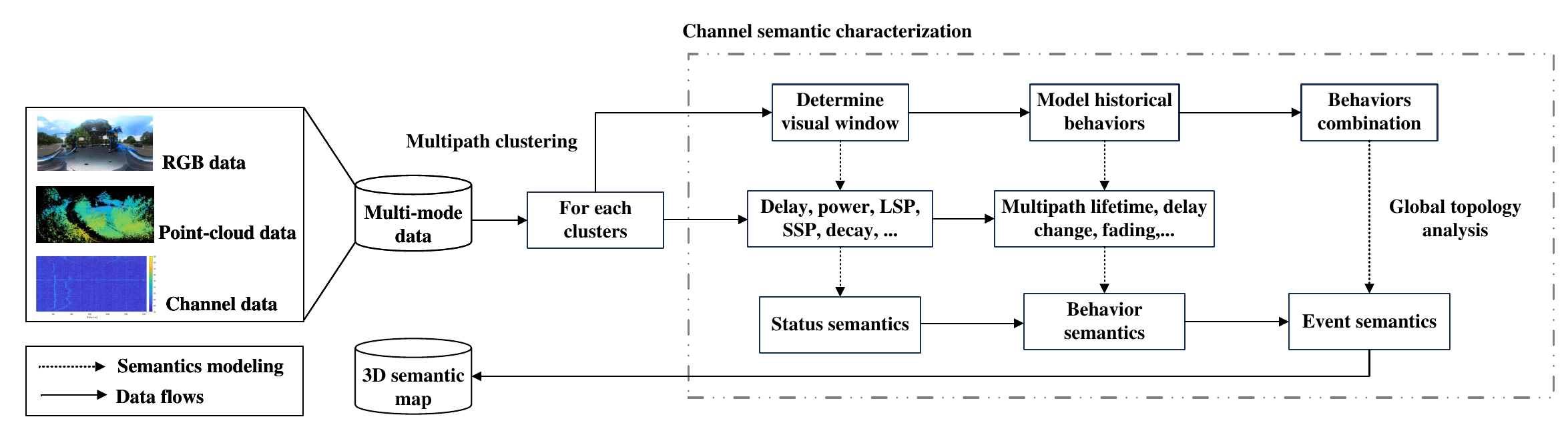}\hspace{0in}
\caption{Channel semantics characterization framework.
\label{fig:model}}\vspace{-0.1in}
\end{figure*}

The channel semantics characterization framework is illustrated as Fig. 3. Through actual channel measurements, we obtained multi-mode channel and environmental data, including RGB data, point-cloud data, and channel data. Power delay profiles (PDPs) are widely used to calculate the received multipaths with propagation delays, which is obtained by the square of the module of CIR as follows:
\begin{equation}
P D P(t, \tau)=|h(t, \tau)|^2
\end{equation}

The muiltipaths in temporal PDP are clustered based on k-power-means algorithms \cite{12}. For each cluster, the intra-cluster parameters, such as delay, power, decay, and other large-scale parameters (LSP) and small-scale parameters (SSP), are calculated. Simultaneously, a visual window on RGB and point-cloud data is determined to associate with this cluster. Specific scatterers are segmented from visual windows, and each cluster is then assigned to a corresponding segmentation zone. Furthermore, channel status semantics are characterized based on cluster parameters derived from channel data and segmentation zones identified in RGB and point-cloud data. During continuous variation periods, time-varying parameters of cluster, such as multipath lifetime, delay changes, fading, and so on, are tracked and extracted. Simultaneously, based on successive RGB data and point-cloud information, historical behaviors of corresponding scatterers are modeled to be associated with this period. Channel behavior semantics are characterized based on time-varying parameters and a historical behavior model, aiming for the same cluster. Combining different behavioral semantics, the whole channel event semantics are characterized based on global topology analysis, thereby deriving a 3D semantic map.

\section{Results and analysis}

\subsection{Channel Status Semantics}

\begin{figure}[htbp]
\centering
\subfigure[]{\includegraphics[width=2.8in]{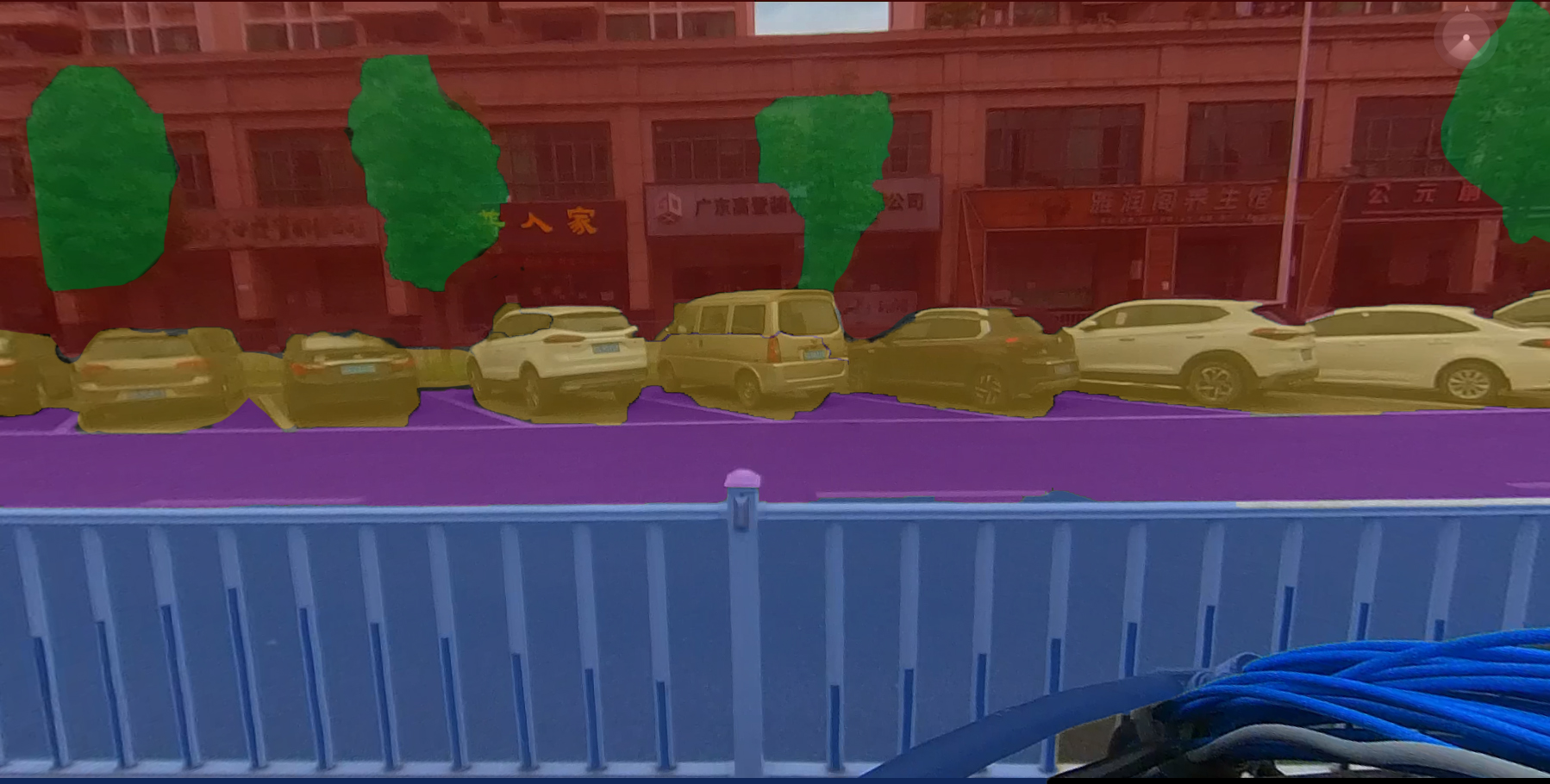}}
\subfigure[]{\includegraphics[width=3in]{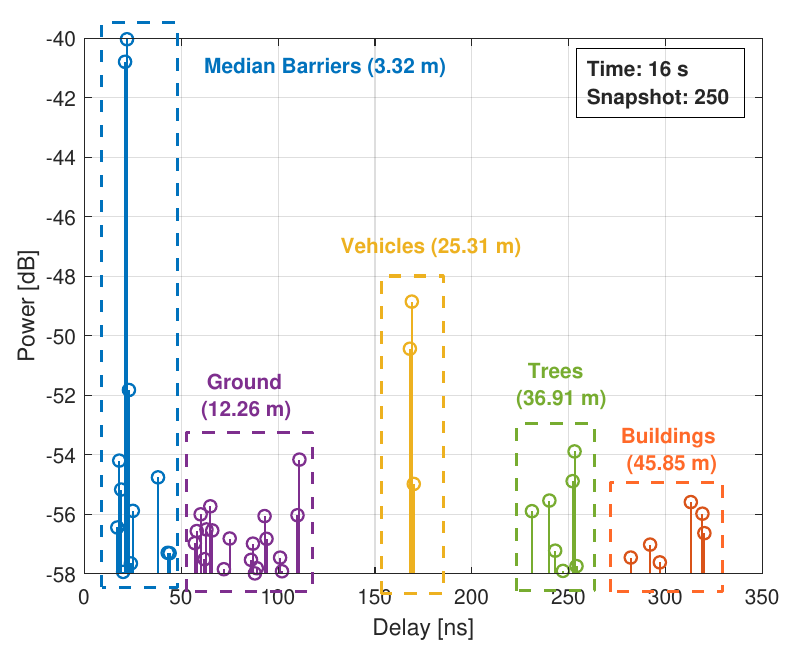}}
\caption{Status semantic characterization results based on actual measurements. (a) Visual window and segmentation zone. (b) Status semantics of clusters}
\label{fig}
\end{figure}

As shown in Fig. 4, the results of channel status semantic characterization based on actual measurements are presented, including visual windows from RGB video and multipaths distribution from channel data. The snapshot index in channel data is 250, corresponding to the RGB data at 16 s. At the current moment, the visual window is semantically segmented into five regions through computer-vision based image segmentation neural networks and manual adjustments, identified as median barriers, ground, vehicles, trees and buildings, as illustrated in Fig. 4(a). Each cluster in the PDP is associated with a semantic segmentation region based on delay and the corresponding propagation distance, resulting in the identification of each cluster. Consequently, the channel status semantics are characterized as median barriers (3.32 m), ground (12.26 m), vehicles (25.31 m), trees (36.91 m) and buildings (45.85 m) respectively. 

From characterization results, channel taps, i.e., multipaths, exhibit distinct features for different channel status semantics. For instance, the power of median barriers is higher than others due to a shorter propagation distance. The power of vehicles comes next, attributed to their strong reflection coefficient. The delay spread of ground, trees, and buildings is greater, accompanied by lower power. The semantic characterization results align with the actual distribution of scatterers in the environment and can be utilized for further extracting status-related channel characteristics and establishing a status semantics library.

\subsection{Channel Behavior Semantics}

\begin{figure}[htbp]
\centering
\subfigure[]{\includegraphics[width=1.4in]{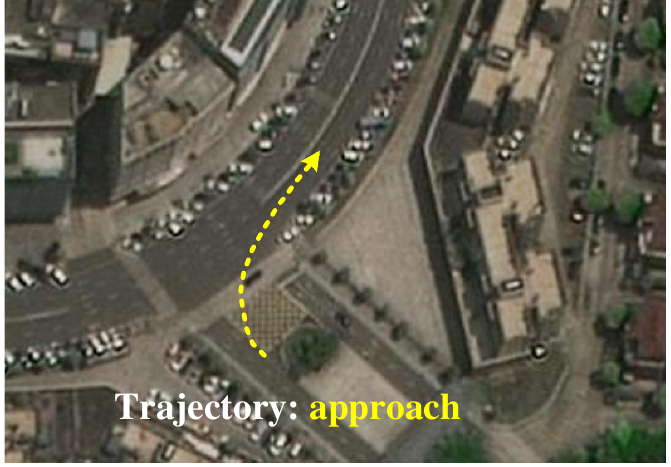}}
\subfigure[]{\includegraphics[width=1.4in]{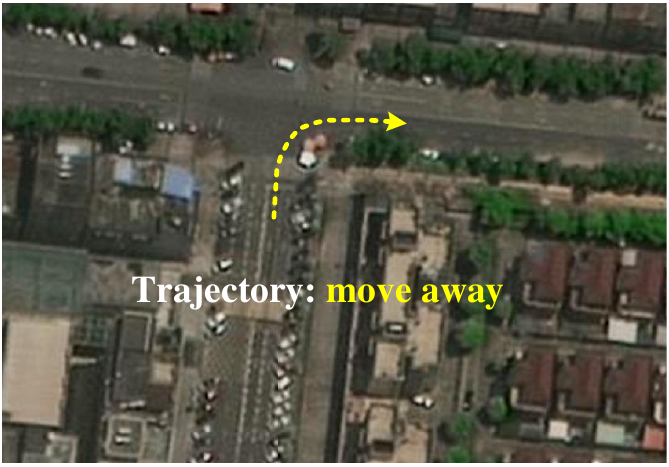}}
\subfigure[]{\includegraphics[width=1.5in]{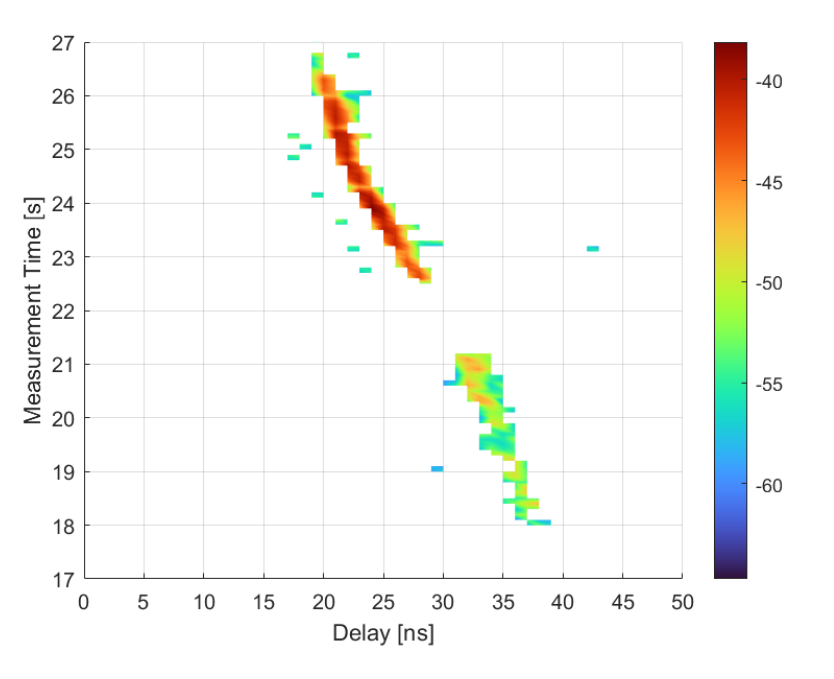}}
\subfigure[]{\includegraphics[width=1.5in]{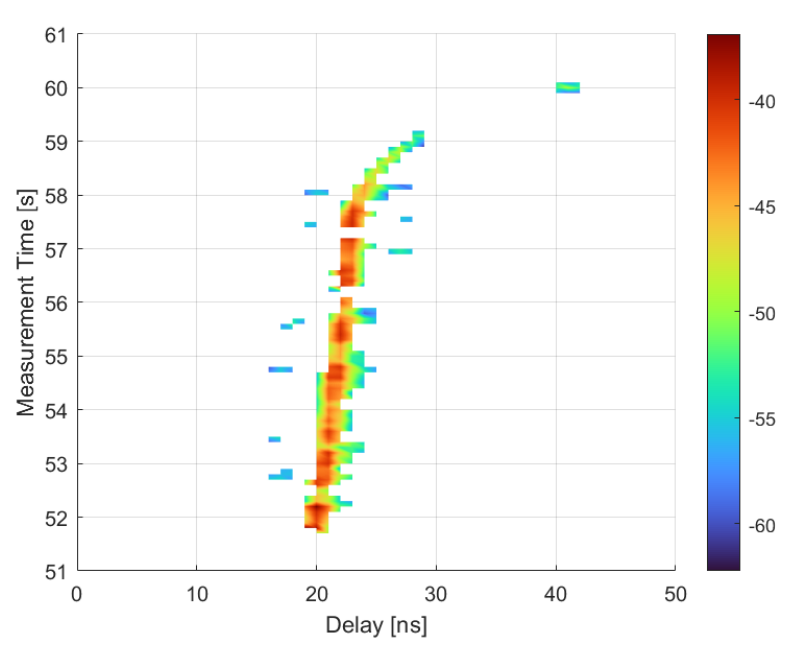}}
\caption{Behavior semantic characterization results based on actual measurements. (a), (b) Historical behaviors. (c), (d) Corresponding time-varying PDPs and trajectory}
\label{fig}
\end{figure}

Fig.5 presents the results of channel behavioral semantics characterization based on actual measurements. Although two typical behaviors are shown in the figure, a variety of distinct historical behaviors can be analyzed in the entire measurement process. For the behavior “approach”, occurring from 17 to 27 s, corresponding to a delay range of 15-40 ns, vehicles are turning to enter a straight road, and the cluster gradually approaches the transceiver. For the behavior “move away”, occurring from 51 to 61 s, corresponding to a delay range of 15-50 ns, vehicles are turning to exit straight road, and the cluster gradually movies away from the transceiver. Fig. 5(c) and 5(d) provides the channel trajectories for these two behaviors, revealing that different behavior semantics exhibit distinct time-varying regulation, such as changes of lifetime, power, and delay. The semantic characterization results align with the actual time evolution and can be utilized for further extracting behavior-related channel characteristics and establishing a behavior semantics library.

\subsection{Channel Event Semantics}

\begin{figure}[!t] \center \vspace{0in}
\includegraphics[width=0.5\textwidth]{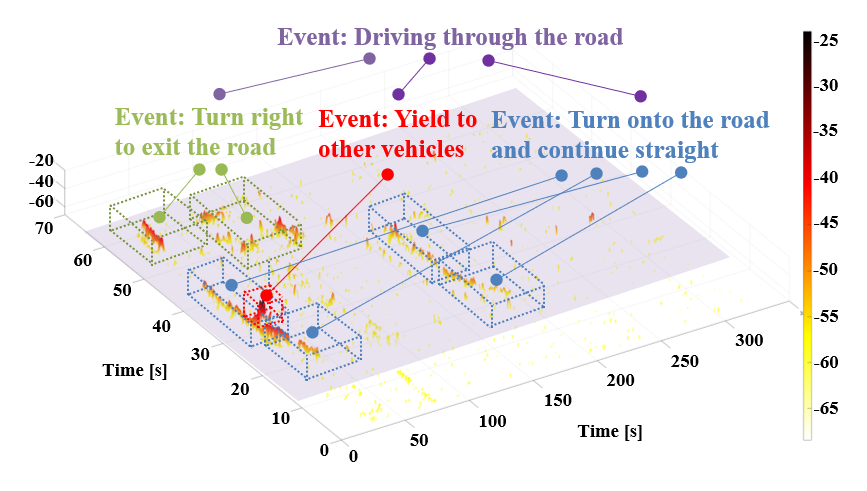}\hspace{0in}
\caption{Event semantic characterization results based on actual measurements.
\label{fig:model}}\vspace{-0.1in}
\end{figure}

Fig. 6 presents the results of channel event semantics characterization based on actual measurements. Different status semantics and behavior semantics combine to low-level event semantics, and these low-level event semantics can further combine to high-level event semantics. For instance, the combination of “median barriers (status semantics) approaching (behavior semantics)” and “vehicles (status semantics) approaching (behavior semantics)” forms “turn onto road (low-level event semantics)”. And “turn onto road and continue straight (low-level event semantics)”, “yield to other vehicles (low-level event semantics)” and “turn right to exit road (low-level event semantics)” further form “driving through road (high-level event semantics)”. The complete channel event semantics exhibit a topology of status, behavior, and event semantics.

\section{Conclusion}

This paper presents characterization of wireless channel semantics. A channel semantic model, including  status semantics, behavior semantics and event semantics, is proposed to associate channel with environment. Based on the actual vehicular channel measurement at 28 GHz, as well as multi-mode data, channel semantic characterization framework and results are provided and analyzed. The research in this paper aims to provide a new paradigm on wireless channel at semantic level, which enable efficient environment perception and understanding for communication system.

\bibliographystyle{IEEEtran}

\bibliography{V2}

\end{document}